\documentclass[%
 superscriptaddress,
 amsmath,amssymb,
 aps, twocolumn,
 pra,
]{revtex4-2}

\usepackage{graphicx}
\usepackage{dcolumn}
\usepackage{bm}
\usepackage{bbm}
\usepackage{amsfonts}
\usepackage{amsmath}
\usepackage{amssymb}
\usepackage{latexsym}
\usepackage{ragged2e}
\usepackage{dsfont}
\usepackage{physics}
\usepackage{mathdots}
\usepackage{color}
\usepackage{gensymb}
\usepackage{caption}
\usepackage{subcaption}

\begin{document}

\preprint{APS/123-QED}

\title{Topological properties of the $[110]$ SnTe nanowires}

\author{Alicja Kawala}

\affiliation{Institute of Theoretical Physics, Jagiellonian University, ulic, S. \L{}ojasiewicza 11, PL-30348 Krak\'ow, Poland}
\affiliation{Jagiellonian University, Doctoral School of Exact and Natural Sciences}

\author{Wojciech Brzezicki}%
 \email{w.brzezicki@uj.edu.pl}
\affiliation{Institute of Theoretical Physics, Jagiellonian University, ulic, S. \L{}ojasiewicza 11, PL-30348 Krak\'ow, Poland}
\affiliation{International Research Centre MagTop, Institute of Physics, Polish Academy of Sciences, Aleja Lotnik\'ow 32/46, PL-02668 Warsaw, Poland}

\date{\today}

\begin{abstract}
SnTe materials are know to be a platform for realization of various strong and symmetry protected topological phases in one, two and three spatial dimensions. We study symmetry-protected topological states in $[110]$ SnTe nanowires in the presence of various combinations of Zeeman field, $s-$wave superconductivity, inversion-symmetry-breaking field and the thickness of the wire. In the normal state we find a Weyl semimetal phase protected by a twofold screw axis symmetry and a topological insulating phase characterized by a $\mathbb{Z}$ invariant protected by a mirror symmetry. 
In the presence of superconductivity, we find inversion-symmetry-protected gapless phase which becomes fully gapped and topologically non-trivial by introducing an inversion-symmetry-breaking field.
Consequently, we find topologically protected localized Majorana zero modes appear at the ends of the wire. We find that this Majorana phase is much easier to achieve in a $[110]$ SnTe nanowire than in a $[001]$ one.
\end{abstract}

\maketitle


\section{Introduction}
SnTe materials (Sn$_{1-x}$Pb$_{x}$Te$_{1-y}$Se$_{y}$) are well known topological crystalline insulators with tunable band inversion based on Sn content \cite{Hsieh2012, Story2012,Tanaka2012,Hasan2012}. Experiments have observed robust 1D modes on surface atomic steps \cite{Sessi2016}, later interpreted as flat bands in a chiral-symmetric model \cite{Buczko2018}. Further studies link these flat bands to a significant zero-bias peak in conductance at low temperatures\cite{Mazur19}, raising speculations about the presence of Majorana modes. However, alternative non-topological explanations exist \cite{Mazur19, Brzezicki2019, timowojtek}. Research also suggests that step defects in 2D SnTe materials could host Majorana end-states \cite{Black-Schaffer, Ojajarvi}.

The improvement in fabricating low-dimensional SnTe systems with tunable electron density is becoming crucial due to theoretical predictions that thin SnTe multilayer systems could support various 2D topological phases, such as quantum spin Hall \cite{Liu2015, Safaei2015}  and 2D topological crystalline insulator phases \cite{Liu2014, Brzezicki2019}. These phases, especially the latter, are of great interest because breaking mirror symmetries in these systems could enable innovative device applications  \cite{Liu2014} and controllable transport properties \cite{kazakov2020dephasing}. It was also shown that twinning boundaries in the SnTe crystals can act like mirror planes that protect crystalline topological order \cite{Buczko_twins}.

Additionally, SnTe materials are promising for studying higher-order topology \cite{Schindlereaat0346,Hsu18, Horizon_penta} and offer a flexible platform for exploring the interplay of different symmetry-breaking fields. 
A study devoted to the $[001]$ SnTe nanowire has shown that it can host hinge and corner states in the normal phase as well as Majorana end-states in the superconducting one  \cite{Brzezicki_snte_nw}. The band gap of the $[001]$ nanowires was also recently studied by DFT methods \cite{Hussain2024} to find inverted gap in the thick wire regime.
It was also shown that the Dirac fermions at the surfaces of the $[001]$ nanowires are free to move around it whereas they remain confined in the $[110]$ nanowires \cite{AB_effect_nste_110}.

Superconductivity in SnTe can be induced via proximity effect or doping  \cite{PhysRevLett.109.217004, PhysRevB.87.140507, Bernevig14, APL17, adma, Bliesener19, PhysRevB.100.241402, trimble2020josephson}, and interesting topological phases emerge in the presence of a Zeeman field \cite{Bernevig14b}, applied via an external magnetic field utilizing huge $g$ factor  $g \sim 50$ \cite{Dybko, leadsalts} or magnetic dopants  \cite{Story86, Story90, PhysRevB.85.045210}. Beyond standard tools of magnetism and superconductivity, SnTe allows for control of topological properties by breaking crystalline symmetries \cite{Chang2016, Kim2019, ZFu2019, Valentine17, Lee2020, Rafal21}, such as inversion symmetry, enabling effects like the giant Rashba effect. 

 Despite a scarce empirical focus on the topological properties of SnTe nanowires, ongoing advancements in fabrication suggest this may soon change \cite{doi:10.1021/acsaelm.0c00740, Sadowski}. Indeed, more experimental groups turn to investigate these materials and learn how to most effectively grow them \cite{PbSnTe_growth}. Recent realizations of pentagonal $[110]$ SnTe nanowires \cite{Horizon_penta} show they would be suitable for studying higher order topology. Furthermore, molecular beam epitaxy is particularly promising because the axial lattice matching between SnTe and GaAs facilitates the growth of high-quality, defect-free heterostructures \cite{Dad2024}. 

 In this paper, we use tight binding model to study symmetries and topological properties of very thin $[101]$ (equivalent to $[110]$) SnTe nanowires. We show a general dependence of phases on the thickness of the nanowire. Throughout this paper we also compare the results with the analysis of $[001]$ SnTe nanowire from \cite{Brzezicki_snte_nw} showing similarities as well as glaring differences.
 
\begin{figure}[!t]
     \centering
     \begin{subfigure}[b]{0.95\columnwidth}
         \centering
\includegraphics[width=0.45\columnwidth]{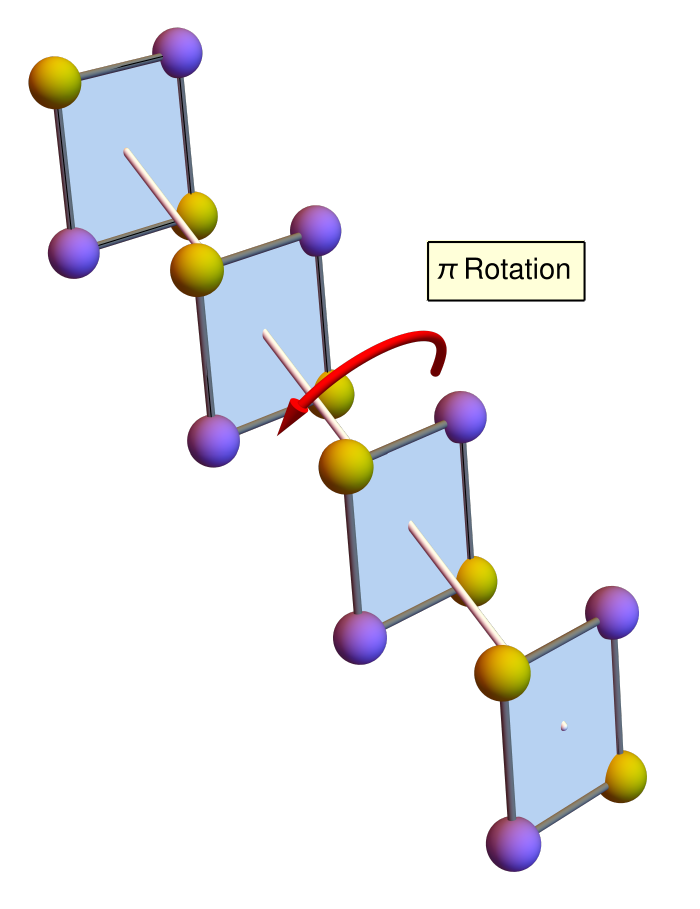}
\includegraphics[width=0.45\columnwidth]{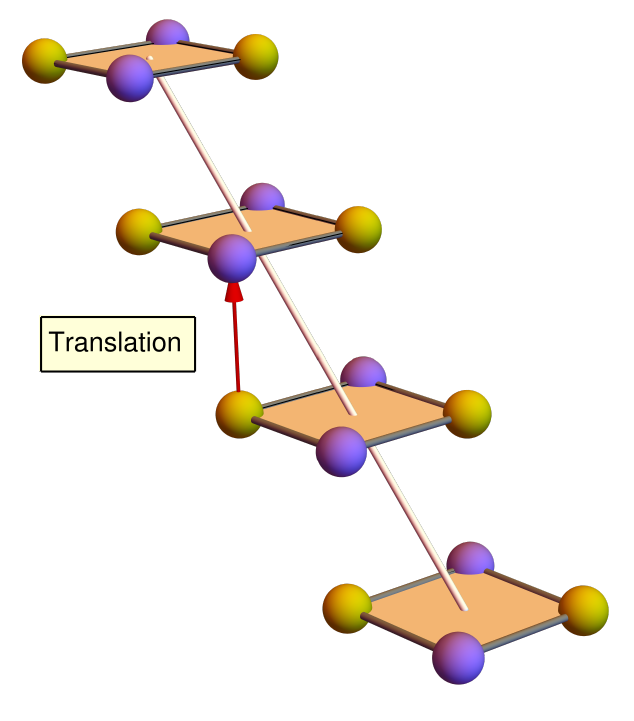}
         \caption{Rotation by $\pi$ and translation by half a unit cell. }
         \label{fig:rot_tr_sym}
     \end{subfigure}
     \vfill
     \begin{subfigure}[b]{0.95\columnwidth}
         \centering
         \includegraphics[width=0.7\columnwidth]{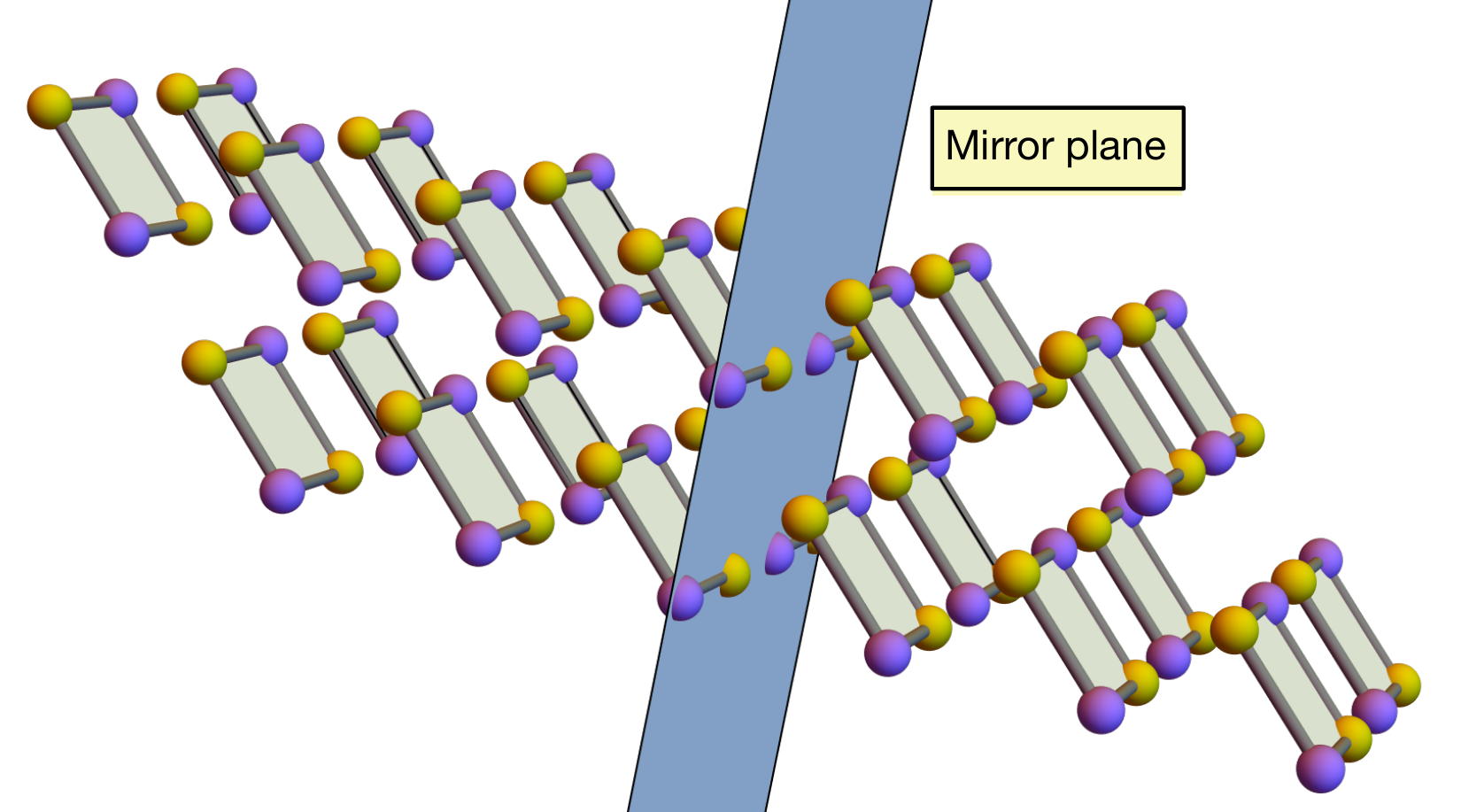}
         \caption{Mirror plane symmetry.}
         \label{fig:mirror_sym}
     \end{subfigure}
\justifying
\caption{\label{fig:symm} Illustration of symmetry operations acting on an infinitely long nanowire. }
\end{figure}

In the Section \ref{sec:model} we describe the tight binding Hamiltonian of the model we adopted. It captures most important physical properties such as spin-orbit coupling and symmetries of the system which are described in the Sec. \ref{sec:symm}. We identify spatial symmetries of the infinite nanowire, i.e. mirror plane symmetry and screw-axis symmetry, Fig. \ref{fig:symm} and discuss possible topological invariants resulting from those symmetries. In Section 2, we introduce a magnetic Zeeman field to the Hamiltonian and analyze the effects of increasing its magnitude and thickness of the nanowire.  We observe alternating insulating and gapless phases, with localized boundary states appearing in the insulating regions. Additionally, we identify an inversion invariant that increases as the system transitions with the growing Zeeman field. In the presence of superconductivity (Section \ref{sec:sup}), we discover inversion-symmetry-protected gapless bulk Majorana modes. Furthermore, introducing an inversion-symmetry-breaking field opens a gap and leads to the emergence of localized Majorana zero modes at the wire's ends. We find that in the case of the $[101]$ SnTe nanowire it is much easier to reach the topologically non-trivial phase than it is for the $[001]$ nanowire.

\section{Tight-binding Hamiltonian\label{sec:model}}
\begin{figure}[!t]
\includegraphics[width=0.45\columnwidth]{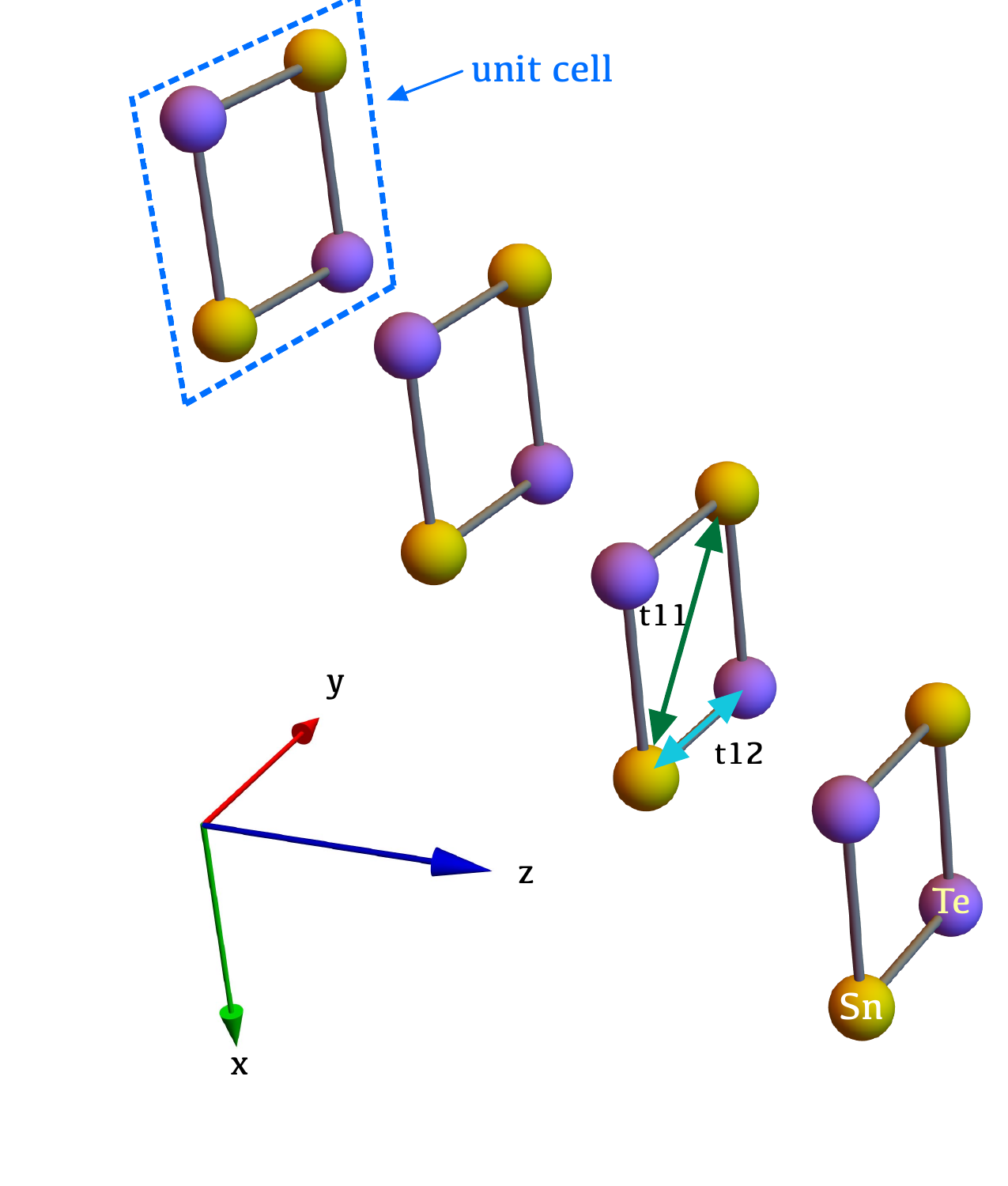}
\includegraphics[width=0.45\columnwidth]{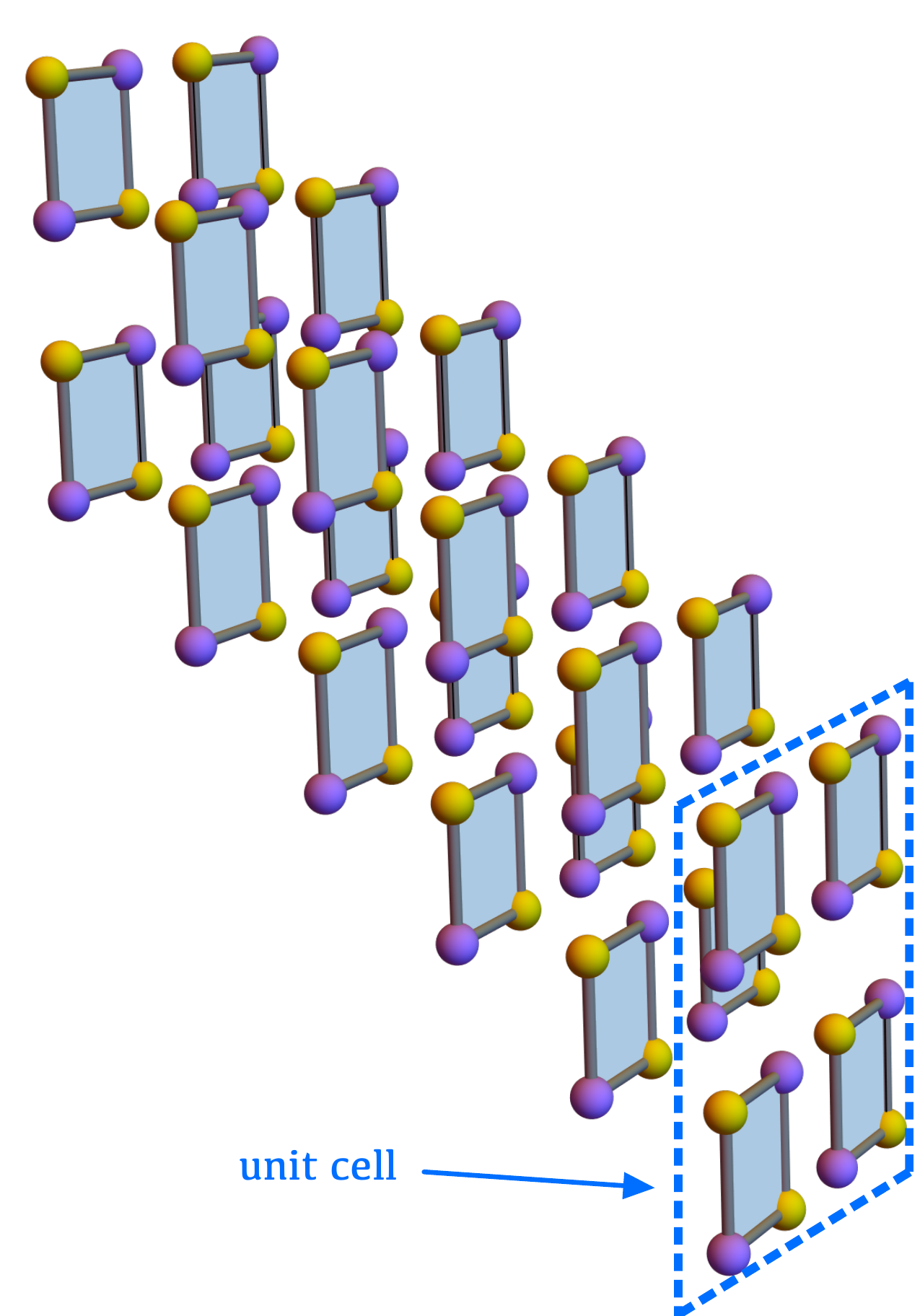}
\justifying
\caption{\label{fig:model}On the left panel, a four-atom unit cell is replicated in a $[101]$ direction (equivalent to $[110]$), establishing the lengthwise span of our model. The thickness of the wire is modified by the repetition of unit cells in the $(001)$ plane. The right image features a wire that is 2 by 2 unit cells thick and 6 unit cells long. }
\end{figure}
A crystal of SnTe has a rock-salt structure with a primitive unit cell that contains one Sn and one Te atom. However, here we use a square-shaped doubled unit cell, see Fig. \ref{fig:model}, to  obtain a desired shape of a nanowire. The lattice translation vectors are
chosen to be $\textbf{a}_1=(2,0,0)$, $\textbf{a}_2=(0,2,0)$ and $\textbf{a}_3=(1,0,1)$, where the distance between neighboring Sn and Te atoms is fixed to be $1$.
To determine electronic properties of the $[101]$ nanowires we use an effective $3p$ tight binding model, which is often used to model SnTe compounds \cite{Hsieh2012,Sessi2016,Brzezicki2019,Brzezicki_snte_nw}. The model contains onsite potential, different for two sublattices of a cubic lattice, nearest-neighbor and next-nearest neighbour hoppings obeying Slater-Koster rules for $p$-orbitals and spin-orbit coupling. It explicit form is given by a Hamiltonian:
\begin{eqnarray}
\mathcal{H}(\textbf{k}) &=&  m \mathds{1}_2\!\otimes\!\mathds{1}_3\!\otimes\!\Sigma + t_{12}\!\!\!\sum_{\alpha=x,y,z}\!\!\mathds{1}_2\!\otimes\!(\mathds{1}_3-\mathit{L}_{\alpha}^{2} )\!\otimes\!h_{\alpha} (\boldsymbol{k}) \nonumber \\
&&\hspace{-0.8cm}+t_{11} \! \sum_{\alpha \neq \beta} \! \mathds{1}_2  \!\otimes\! \left[ \mathds{1}_3 - \dfrac{1}{2}\left(\mathit{L}_{\alpha}+ \epsilon_{\alpha\beta}\mathit{L}_{\beta}\right)^2\right] \!\otimes\! h_{\alpha,\beta} (\boldsymbol{k}) \Sigma  \nonumber \\
&&\hspace{-0.8cm}+ \lambda\!\! \sum_{\alpha=x,y,z} \!\! \,\sigma_{\alpha} \!\otimes\! \mathit{L}_{\alpha} \!\otimes\! \mathds{1}_4,
\label{eq:ham}
\end{eqnarray}
where $\epsilon_{\alpha\beta}$ is a Levi-Civita symbol, $\mathds{1}_n$ are $n \times n$ identity matrices, $L_\alpha$ are $3 \times 3$ angular momentum $L=1$ matrices acting in the basis of cubic orbitals $p_x$, $p_y$ and $p_z$, $\sigma_\alpha$ are Pauli matrices,  $\Sigma$ is a diagonal $4\times 4$ matrix with entries $\pm 1$ specifying which lattice site is occupied with which atom (Sn or Te). The matrices $h_\alpha(\boldsymbol{k})$, $h_{\alpha, \beta}(\boldsymbol{k})$ indicate hoppings between nearest and next-nearest neighbours, given explicitly in Appendix \ref{app:nanowire}. We use following values of the hopping amplitudes between Sn-Te atoms: $t_{12} = 0.9\textrm{ eV} $, and between Sn-Sn, Te-Te atoms: $t_{11} = t_{22} = 0.5\textrm{ eV} $. A spin-orbit coupling constant is set to $\lambda = 0.3\textrm{ eV}$. These values were already used in earlier works on SnTe materials \cite{Brzezicki2019,Brzezicki_snte_nw}.

Hamiltonian (\ref{eq:ham}) describes a three-dimensional bulk SnTe. To get the Hamiltonian of a $[101]$ nanowire (NW) we need to impose open boundary conditions along lattice direction $\textbf{a}_1$ and $\textbf{a}_2$ and fix the number of unit cells stacked in these directions. This transformation fixes the width of a nanowire.
By keeping boundary condition periodic along $\textbf{a}_3$ we are able to study a bulk NW, i.e., NW which has no ends. We describe this system by a Hamiltonian denoted as  
$\mathcal{H}^{\rm NW}(k_3)$. Its construction is described in details in Appendix \ref{app:nanowire}.

\section{Symmetries\label{sec:symm}}

Symmetries of the bulk SnTe are well known and were discussed, e.g., in \cite{Brzezicki2019,Brzezicki_snte_nw}. However, symmetries of the  $[101]$ NW are somewhat different due to the large unit cell and open edges at the sides of the NW. What is most specific to this geometry (see Fig. \ref{fig:model}) is presence of a twofold screw axis, i.e., a nonsymmorphic symmetry that consists of a $\pi$-rotation with respect the $[101]$ axis accompanied by a half lattice vector translation along the same axis (see Fig. \ref{fig:symm} for a schematic picture). 

On the algebraic level the symmetry is represented by an operator $S_c(k_3)$, where $k_3$ is the quasimomentum along the NW. The exact form of $S_c(k_3)$ is given in the Appendix \ref{app:nanowire}.

Screw axis symmetry commutes with the NW Hamiltonian for every $k_3$,
\begin{equation}\label{eq:comm}
[\mathcal{H}^{\rm NW}(k_3),S_c\left(k_{3} \right)]\equiv 0.
\end{equation}
 Obviously,  $S_c(k_3)$ is an order-two symmetry, i.e. $S_c(k_3)^2\equiv 1$. We also find that ${\rm Tr}S_c(k_3)\equiv 0$, so it needs to have equal number of $+1$ and $-1$ eigenvalues. A block decomposition of $\mathcal{H}^{\rm NW}(k_3)$ in the eigenbasis of $S_c(k_3)$ follows from the commutation rule \ref{eq:comm}: 
\begin{equation}
\tilde{\mathcal{H}}^{\rm NW}(k_3) = \mathcal{H}_+(k_3)\oplus\mathcal{H}_-(k_3),
\end{equation}
where indices $\pm$ indicate eigenvalues of $S_c(k_3)$. 

The nonsymmorphic character of $S_c(k_3)$ results in the doubling of the period of the $\mathcal{H}_{\pm}(k_3)$. This means that $\mathcal{H}_{\pm}(k_3)\not=\mathcal{H}_{\pm}(k_3+2\pi)$ but $\mathcal{H}_{\pm}(k_3)=\mathcal{H}_{\pm}(k_3+4\pi)$.

Apart from screw axis the NW also has an inversion symmetry $I$ that satisfies
usual relation
\begin{equation}
I\mathcal{H}^{\rm NW}(k_3)I^{-1}\equiv \mathcal{H}^{\rm NW}(-k_3).
\end{equation}
This symmetry is also inherited by the diagonal blocks $\mathcal{H}_{\pm}(k_3)$, so they are both inversion-symmetric - see Appendix \ref{app:nanowire} for details.

In the absence of magnetic field the NW Hamiltonian has a usual spinful time-reversal symmetry $\cal T$ satisfying
\begin{equation}
{\cal T}\mathcal{H}^{\rm NW}(k_3){\cal T}^{-1}\equiv \mathcal{H}^{\rm NW}(-k_3),
\end{equation}
and ${\cal T}^2=-1$. Since there are no other symmetries present we can conclude that the system is in the AII Altland-Zirnbauer class in the absence of magnetic field and in the A class in its presence. On the other hand, if we consider $\mathcal{H}^{\rm NW}(k_3)$ in the screw-axis eigenbasis we find that $\mathcal{H}_{\pm}(k_3)$ blocks have no time-reversal symmetry even in absence of the magnetic field - see Appendix \ref{app:nanowire}.

Concerning possible topological invariants of the system in an insulating phase: from the standard tenfold \cite{PhysRevB.55.1142} classification in one spatial dimension it follows that  $\mathcal{H}(k_3)$ is trivial both with and without magnetic field. If we also take into account that the system has a nonsymmorphic order-two symmetry that flips zero coordinates of quasimomentum then from the work of Sato and collaborators \cite{Sato_NS} we find that it is still trivial.

Looking for a non-vanishing invariant we can think of inversion symmetry $I$. From Ref. \cite{Aris_inv} we can learn that it is a non-negative integer ($\mathbb{Z}^{\ge}$) denoted as $N_{(-1)}$. It can be calculated as the difference of number (neglecting the sign) of occupied bands with $-1$ inversion eigenvalue between $k_3=0$ and $k_3=\pi$. 

Another option of having a non-vanishing invariant is more exotic. Combining together inversion and screw axis we get a relation with the Hamiltonian in a form of: 
\begin{equation}
[IS_c(k_3)]\mathcal{H}^{\rm NW}(k_3)[IS_c(k_3)]^{-1}\equiv \mathcal{H}^{\rm NW}(-k_3).
\end{equation}
From the geometric point of view by applying $IS_c(k_3)$ we combine inversion with $\pi$-rotation and half lattice vector translation. 
The result is a $[101]$-mirror reflection accompanied by a half lattice vector translation along $[101]$ axis. 
Since the translation is perpendicular to the mirror 
plane it does not lead to a nonsymmorphic symmetry but rather to
a regular mirror reflection depicted in Fig. \ref{fig:symm}. 
Therefore we denote this symmetry as:
\begin{equation}
M_r(k_3)\propto IS_c(k_3),
\end{equation}
where the dependence on $k_3$ follows from the unit cell on the NW being incompatible with the symmetry. Although it is impossible to choose such a unit cell in case of the present NW one can at least choose it to lie on the mirror plane, see Appendix \ref{app:mirror} for details. Concerning topology,
according to the general classification of Ref. \cite{Chiu_reflection} the system can host
non-trivial $M\mathbb{Z}$ invariant in presence or absence of longitudinal Zeeman magnetic field. 

\begin{figure}
\includegraphics[width=1.0\columnwidth]{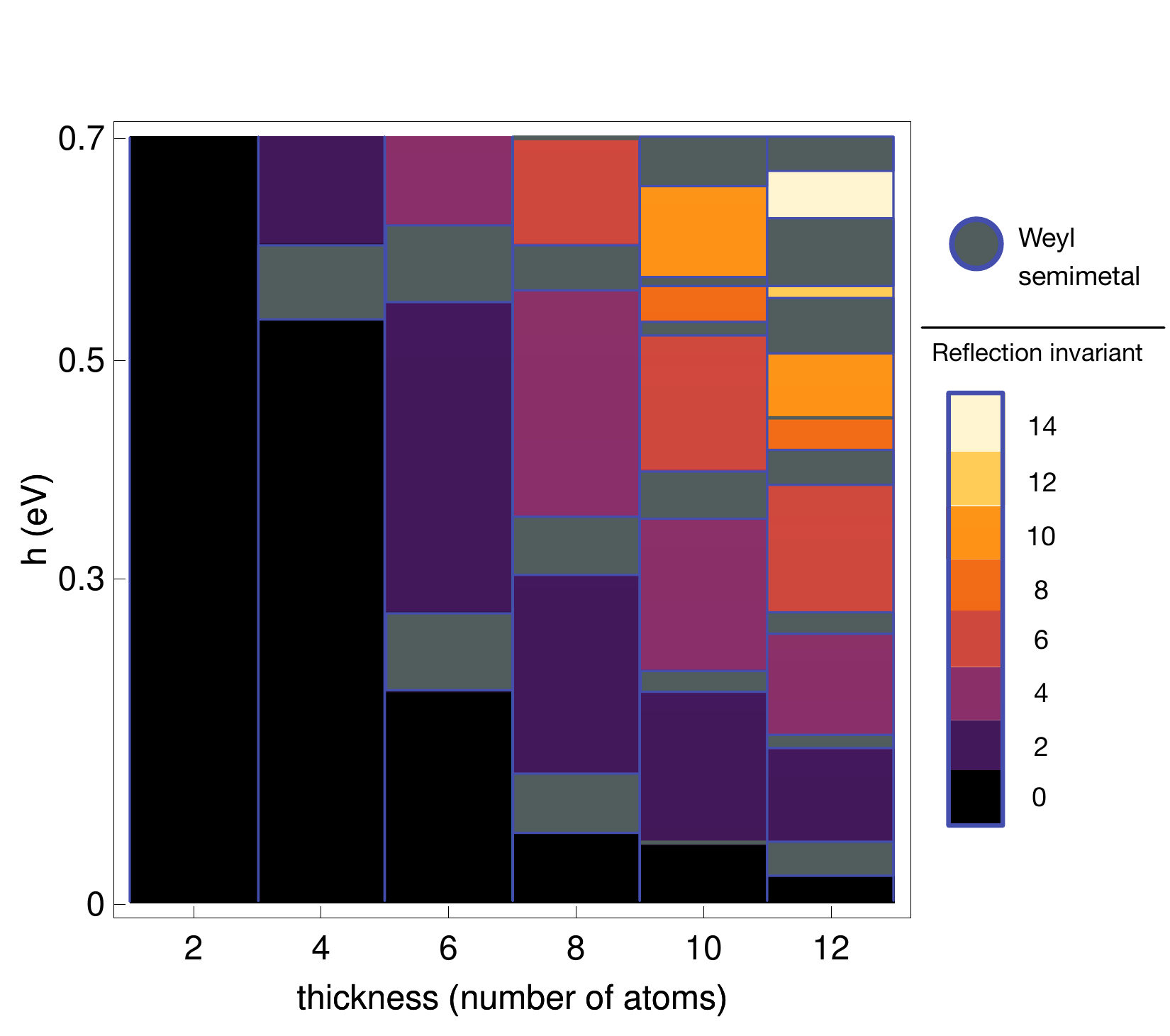}
\caption{\label{fig:pd} Phase diagram of the system as a function of the Zeeman field's magnitude $h$ along the nanowire and the thickness measured in number of atoms. Color map regions specified by the reflection invariant denote insulating phase. Black regions with zero invariant have simple insulator band structure, whereas higher invariant regions demonstrate gap-inverted band structure hosting ingap boundary states. The semimetallic phase where gap closes is marked by a grey colour. 
}
\end{figure}

\begin{figure*}[!t]
\includegraphics[width=0.99\textwidth]{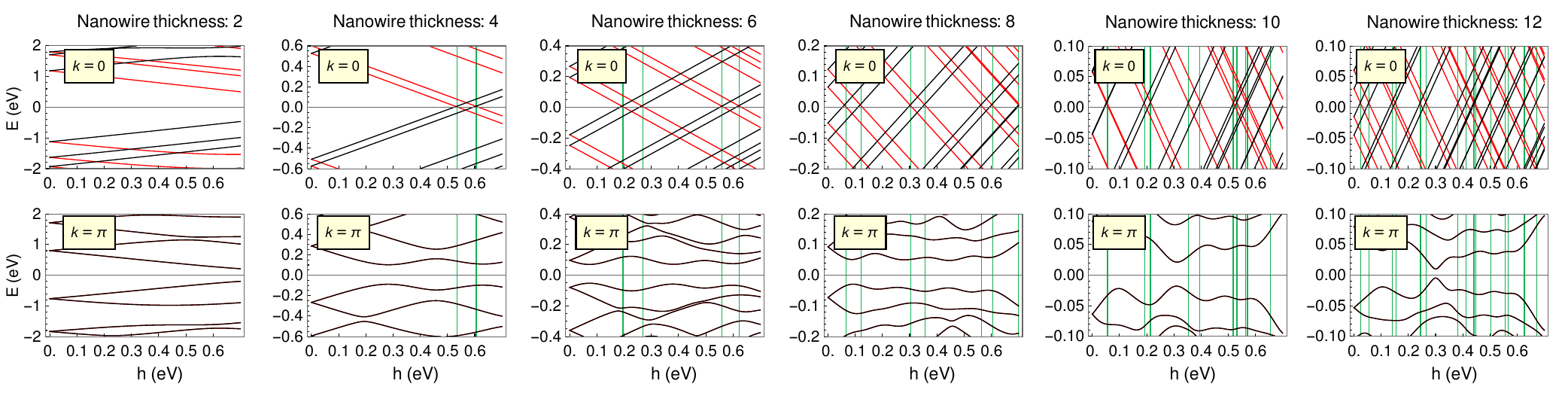}
\caption{\label{fig:refl}  
Energy bands of the Hamiltonian $\mathcal{H}^{\rm NW}_{\rm mag}(k)$ as a function of the Zeeman field's magnitude $h$ for $k=0$ (top row) and $k=\pi$ (bottom row). The bands are colored according to the value of the reflection operator, red color corresponding to value -1, and black to value +1. Green vertical grids mark the magnetization steps from the Fig. \ref{fig:magn}.
}
\end{figure*}

\section{Topological properties of the normal state}\label{s:zeeman}

\justifying
We introduce a magnetic field to the Hamiltonian (\ref{eq:ham}) in the form of Zeeman coupling in the longitudinal direction of the wire:
\begin{equation}
\mathcal{H}^{\rm NW}_{\rm mag}(k_3) = \mathcal{H}^{\rm NW}(k_3)+\mathds{1}_{\!N_x}\otimes\mathds{1}_{\!N_y}\otimes
    \left( \vec{h}\cdot \vec{\sigma}\right) \otimes\mathds{1}_3\otimes \mathds{1}_4,
\end{equation}
where $\vec{\sigma}$ is a vector of Pauli matrices and $\vec{h}$ indicates the direction and magnitude of the field. In our case $h = |\vec{h}|$ and $\vec{h}$ is pointing along the wire. The resulting phase diagram as function of $h$ and the thickness of the wire (assuming square section) is given in Fig. \ref{fig:pd} and more details on calculation of the gap are given in Appendix \ref{app:gap}. For the thinnest wire we observe only a trivial insulating phase within the range of considered here magnetic field. By increasing the thickness we start seeing a cascade of phases with increasing reflection invariant $M\mathbb{Z}$ as the field is increased. The invariant grows monotonically with the field in steps of height $2$ and different topological phases are separated by the gapless ones. These are Weyl semimetal phases where the gap closes due to crossing of bands belonging to different screw-axis symmetry subspaces.

The values of the  $M\mathbb{Z}$ invariant can be inferred from the energy spectrum of the model at high-symmetry points. In Fig. \ref{fig:refl} we show the spectra at $k_3=0$ and $k_3=\pi$ for increasing values of the Zeeman field $h$ colored according to the eigenvalue of the mirror-symmetry operator $M_r(k_3)$.
We see similar behavior for different values of the NW thickness: the energy gap never closes at $k_3=\pi$ and there are multiple gap closings at $k_3=0$. The gap closing is always realized as crossing of $M_r=-1$ energy level with positive velocity with $M_r=+1$ energy level with negative velocity. Close to the Fermi energy at $k_3=0$ all energy levels are linear in $h$ and have the same slope for a given $M_r$ eigenvalue, being opposite for $M_r=\pm 1$. It is clear that the $M\mathbb{Z}$ invariant can be formulated as number of occupied $M_r=+1$ energy levels and we see that it changes by even number between different insulating phases. 

\begin{figure}[h]
\includegraphics[width=0.99\columnwidth]{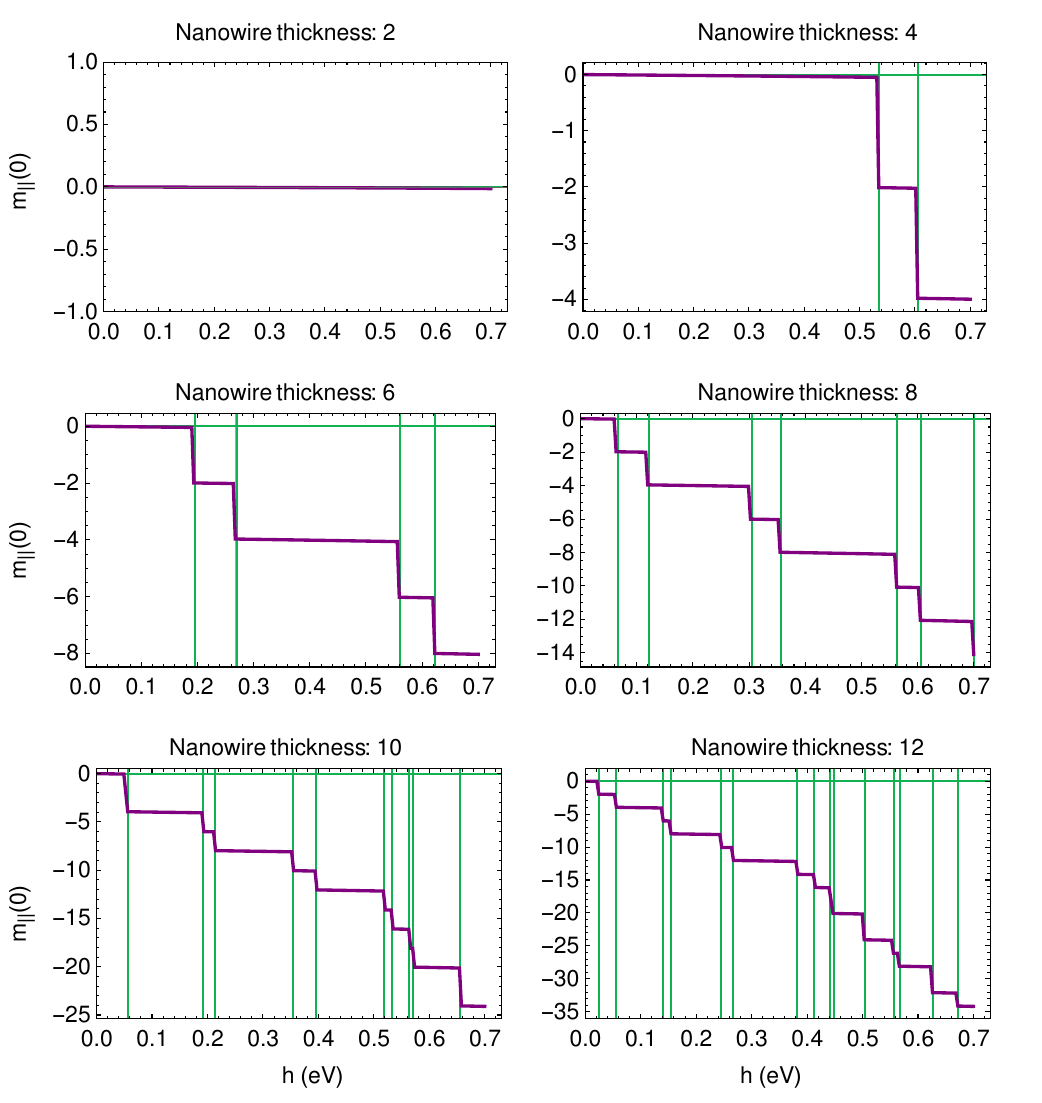}
\caption{\label{fig:magn} 
Magnetization contribution from bands at $k=0$ as a function of the Zeeman's field magnitude $h$. Vertical green lines mark discontinuous steps.}
\end{figure}

Quite remarkably the change in topology of the insulating states can be associated with the behavior of the longitudinal magnetization of the NW. It is best seen if we look at it as $k$-resolved quantity. We define is as 
\begin{eqnarray}
m_{\parallel}(k_3) &=& \sum_{n=1}^{n_f} \langle E_n(k_3)|\,\mathds{1}_{\!N_x}\otimes\mathds{1}_{\!N_y}\otimes
    \left( \sigma_x+\sigma_z\right)/\sqrt{2} \nonumber\\
    &\otimes&\mathds{1}_3\otimes \mathds{1}_4 \,|E_n(k_3)\rangle
\end{eqnarray}
where $n_f=12N_xN_y$ indicates Fermi level and $|E_n(k_3)\rangle$ are the eigenstates of $\mathcal{H}^{\rm NW}_{\rm mag}(k_3)$. The behavior of $m_{\parallel}(0)$ as function of $h$ is shown in Fig. \ref{fig:magn}. We see that the magnetization at $k_3=0$ changes in steps of two at the transition points between different $M\mathbb{Z}$-phases. This behavior is related to the fact that within the spin sector the mirror symmetry $M_r(0)$ that yields the $M\mathbb{Z}$ invariant is indeed identical to $m_{\parallel}(0)$ however it is not clear why other sectors (orbital, intra-cell and inter-cell) do not seem to play any role at the Fermi surface.     

Note that similar cascade of insulating phases was previously found for a $[001]$ SnTe NW studied in Ref. \cite{Brzezicki_snte_nw}, however in this former case it was not possible to assign any bulk topological invariant to them. Consequently the topology of $[001]$ NWs was characterized by sublattice pseudospin textures. In the present case it is also possible to notice such non-trivial textures in the topological phases, see Appendix \ref{app:stext}, however we argue that characterization by the $M\mathbb{Z}$ is more accurate.    

\begin{figure}[!t]
\includegraphics[width=0.95\columnwidth]{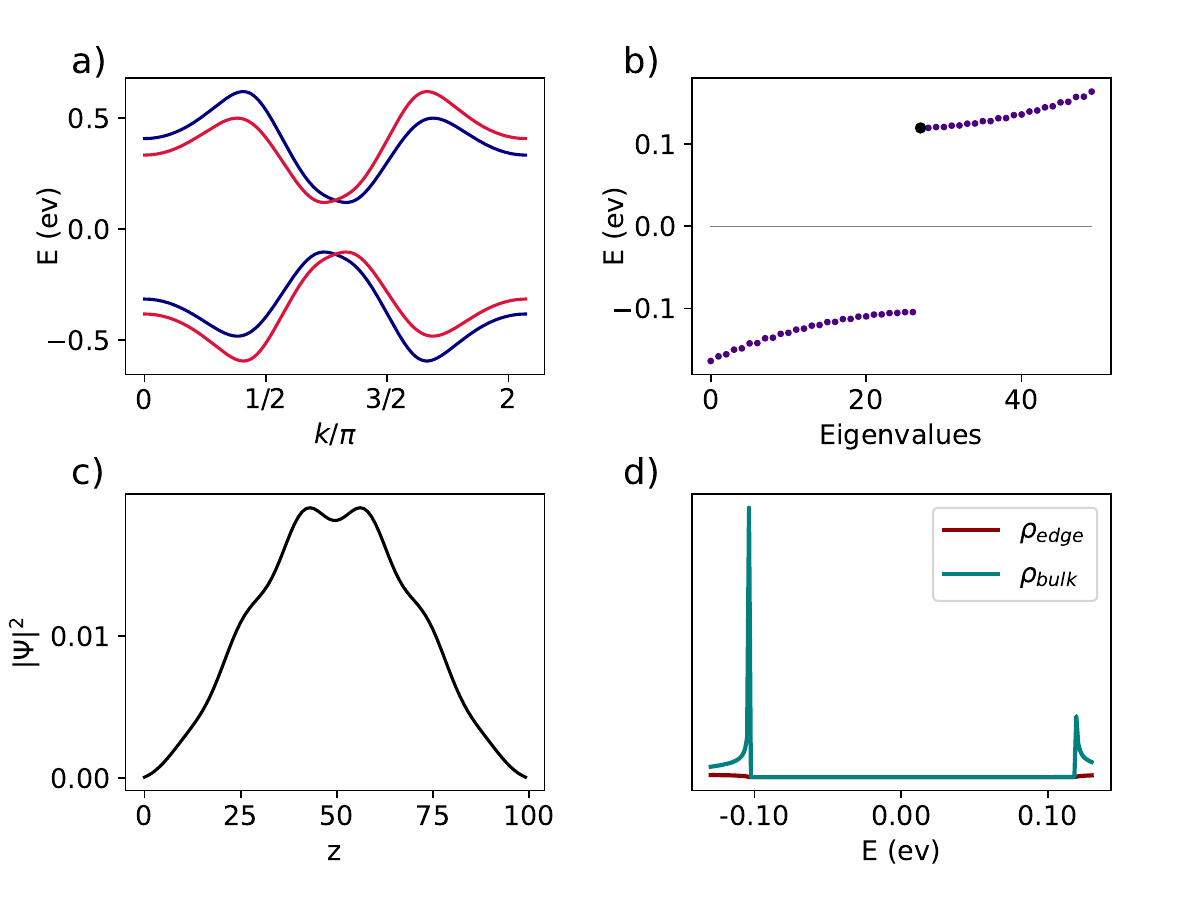}
\caption{\label{fig:ins} Electronic structure of the trivial insulating nanowire (4$\times$4 atoms thick  in 0.2 eV Zeeman field). Electronic bands of two independent parts of block-diagonalized Hamiltonian - a). Eigenvalues of the fully open wire - b). Amplitudes of the eigenvector to the eigenvalue marked by black dot on plot b) ($z$ is indexing the layers arranged lengthwise) - c). Density of states in the bulk, $\rho_{bulk}$, and at the ends of the wire, $\rho_{edge}$ - d). }
\end{figure}

Having established bulk properties of the insulating phases we can now look at open nanowires to search for end-states. In Fig. \ref{fig:ins} we show the case of trivial insulator for $4$-atom thick NW. The band strucure around the gap is shown in Fig. \ref{fig:ins}(a) where different colors represent different screw axis symmetry subspaces.
We see two pairs of parabolic bands with gap around $k_3=\pi$.
Small splitting of energies between different $S_c$ sectors is caused  by the Zeeman field which vanishes at $k_3 = \pi$ due to the nonsymmorphic character of screw axis symmetry.
The energy spectrum for finite open system is given in Fig. \ref{fig:ins}(b) and it exhibits no in-gap states, as expected. For completeness we show the local density of states for the first state above the gap in plot \ref{fig:ins}(c) and the densities of states for the bulk $\rho_{bulk}$ and for the edge $\rho_{edge}$ in a semi-infinite system - see Fig. \ref{fig:ins}(d). For the latter plot we have used the iterative Green's function method given in Ref. \cite{Sancho_1985}. Analogical plots are done in the non-trivial phase of a $6$-atom thick NW with the $M\mathbb{Z}$ invariant being equal to $2$.  
\begin{figure}[!t]
\includegraphics[width=0.95\columnwidth]{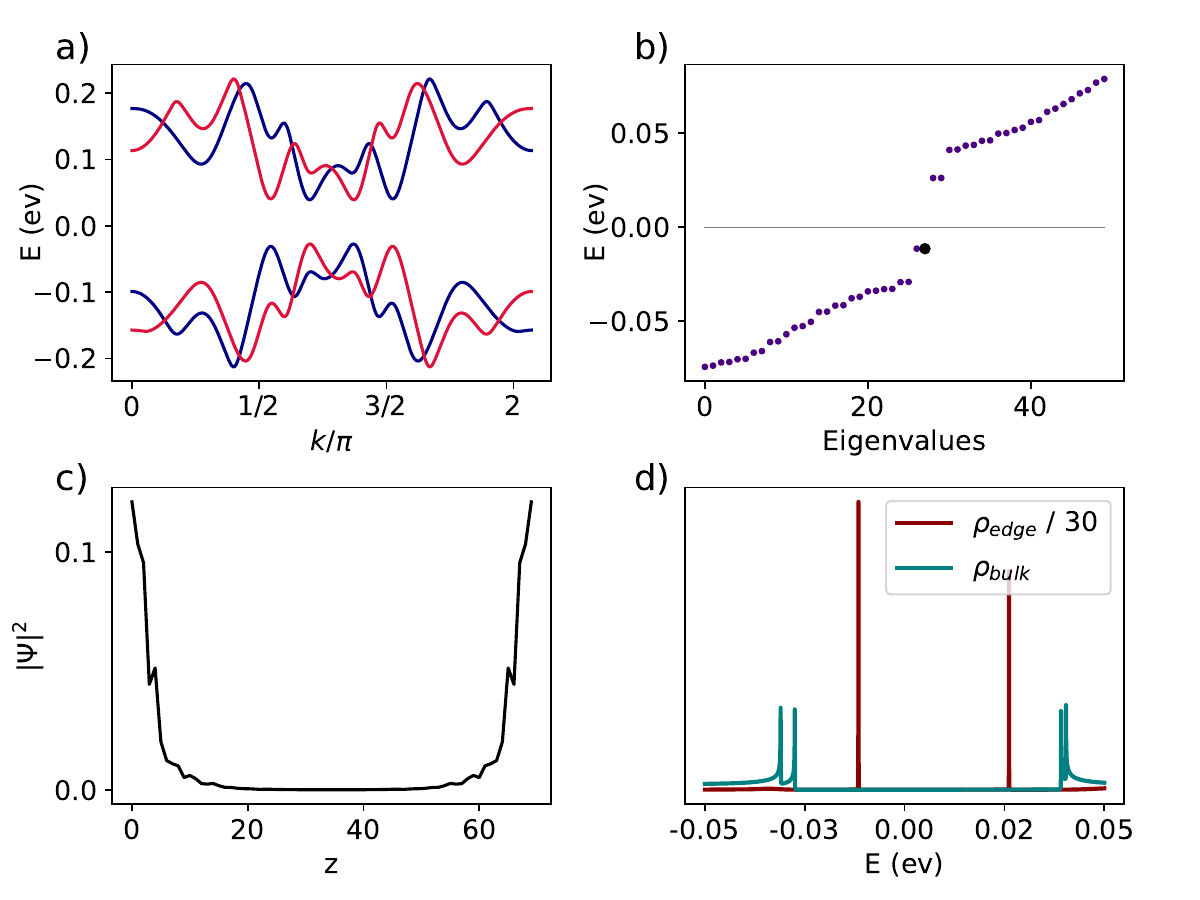}
\caption{\label{fig:tins}Electronic structure of the band-inverted insulating nanowire (3$\times$3 unit cells thick in 0.45 eV Zeeman field). Electronic bands of two independent parts of the block-diagonalized Hamiltonian - a). Eigenvalues of the fully open wire - b). Amplitudes of the eigenvector to the eigenvalue marked by black dot on plot b) ($z$ is numbering the layers arranged lengthwise) - c). Density of states in the bulk, $\rho_{bulk}$, and at the ends of the wire, $\rho_{edge}$ - d). }
\end{figure}
\begin{figure}[!t]
\includegraphics[width=0.95\columnwidth]{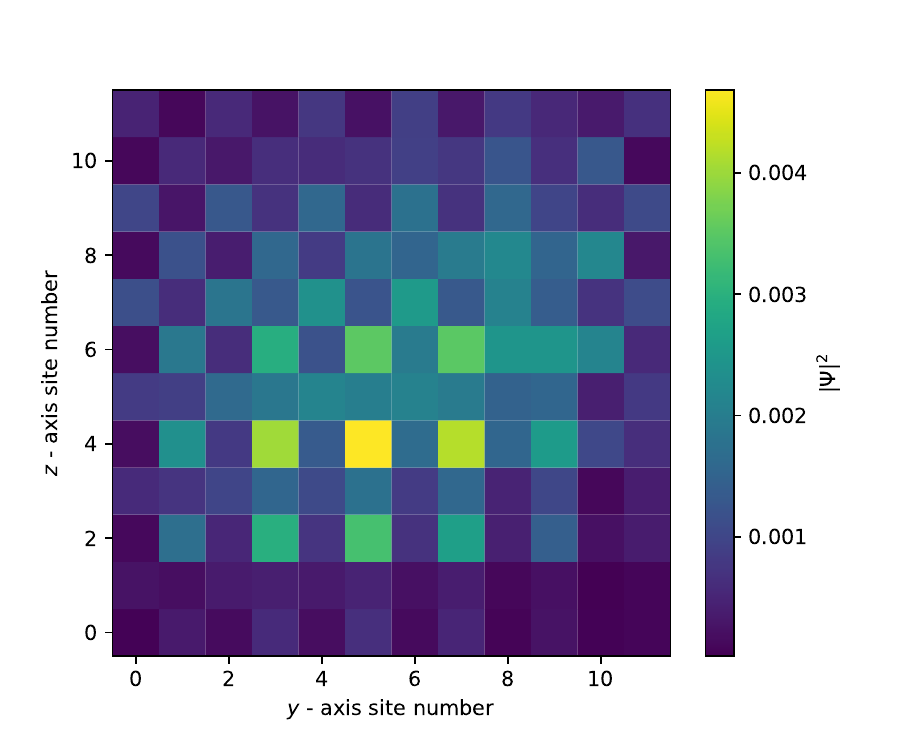}
\caption{\label{fig:denins} Local density of states for a pair of end-states with energy E = -0.0073 eV,  projected onto first layer of the nanowire for the case of 12 atoms thick wire in an insulating phase.
}
\end{figure}
We can see band structure having multiple minima around $k_3=\pi$, see Fig. \ref{fig:tins}(a), and energy spectrum for open system exhibits two pairs of in-gap states, as shown in Fig. \ref{fig:tins}(b). The plot \ref{fig:tins}(c) of local density of states for one of these end-states clearly shows localization at the ends of the NW. This is confirmed by the calculation of  $\rho_{bulk}$ and $\rho_{edge}$ done for semi-infinite system. It is also interesting to look at the location of the end-states in the transverse cross-section of the NW. In the case of $[001]$ NW studied before they had clear tendency to appear in the corners of the NW. This is not the case here. In Fig. \ref{fig:denins} we show local density of states for a pair of equal-energy end-states found in $12$-atom thick NW. The spectral weight is clearly concentrated around the center of the NW and there are no signatures of corner states. We also see that this density profile has no spatial symmetry which follows from the fact that the termination of the NW breaks them all. It is worth emphasizing here that any termination of the $[101]$ NW will break the $M_r$ mirror symmetry therefore the end-states are generically not guaranteed to appear within the gap. 

\begin{figure}[!t]
\includegraphics[width=0.95\columnwidth]{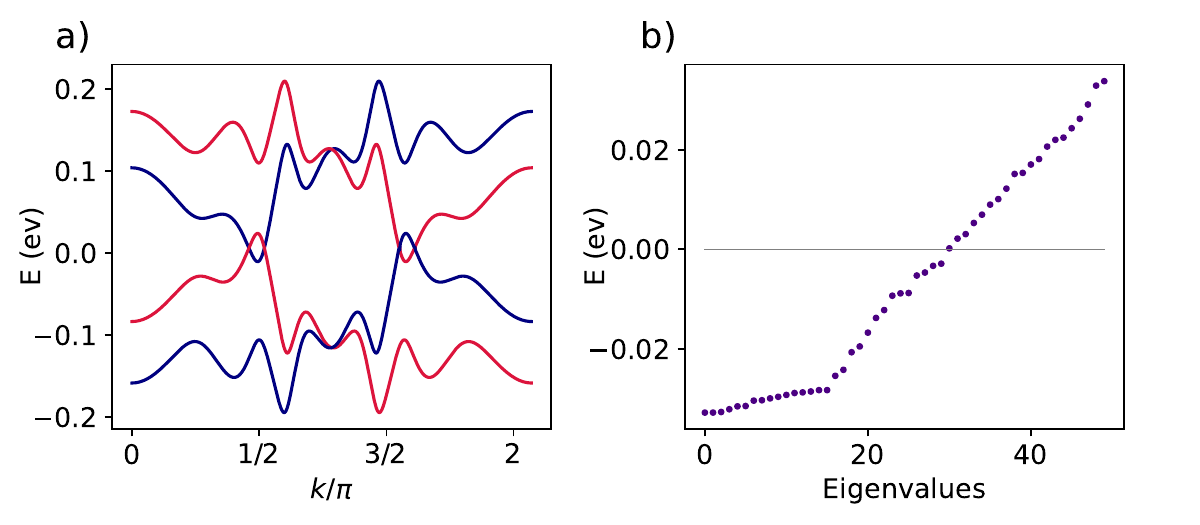}
\caption{\label{fig:metal} Electronic structure of the semi-metal phase 2$\times$2 unit cells thick naowire in 0.7 eV Zeeman field. Electronic bands of two independent parts of the block-diagonalized Hamiltonian - a). Eigenvalues of the fully open wire - b).
}
\end{figure}

Finally, in phase diagram of Fig. \ref{fig:pd} we also find gapless phases. These are semimetal phases where bands from different screw-axis symmetry subspaces cross. Since the dispersion is linear around gap closing points and spin degeneracy is lifted we call then Weyl semimetal phases. The exemplary band structure around the Fermi level and spectrum of an open NW is shown in Fig. \ref{fig:metal}. We can see two pairs of Weyl points with opposite topological charges. For Weyl point located at $k_3=k_0$ such a charge can be defined as a difference of number of occupied states with $S_c=+1$ eigenvalue at $k_3=k_0-\epsilon$ and $k_3=k_0+\epsilon$. Obviously, in one spatial dimension gapless phase cannot give any end-state in an open system.

\section{Superconducting phase\label{sec:sup}}

The realization of Majorana modes at the ends of the superconducting nanowire is a long sought-after phenomena which could potentially revolutionize quantum computing. A realistically achievable model was proposed in \cite{Lut10} as a semiconductor with superconductivity induced by the proximity effect. We analyze the SnTe NW in the case of the induced s-wave superconductivity. The Hamiltonian needs to be appropriately modified to include particle-hole degree of freedom and pairing amplitude. It can be accomplished using the following formula:
\begin{equation}\label{eq:sc}
\mathcal{H}^{\rm NW}_{\rm sc}(k_3) = \begin{pmatrix} \mathcal{H}^{\rm NW}_{\rm mag}(k_3) - \mu & i\sigma_{y}\Delta \\
-i\sigma_{y}\Delta & -(\mathcal{H}^{\rm NW}_{\rm mag}(-k_3)^T - \mu) \\
\end{pmatrix},
\end{equation}
where  $\sigma_{y}$ is the Pauli operator acting on the spin degrees of freedom; $\Delta$ and $\mu$ are the magnitude of the superconducting gap and the chemical potential at which the gap opens. The superconducting Hamiltonian $\mathcal{H}^{\rm NW}_{\rm sc}(k_3) $ gains particle-hole symmetry in the form of:
\begin{equation}
    \mathcal{C} = C \cal K,
\end{equation}
with $\cal K$ being complex conjugation and $C$ being a unitary operator given by a block matrix:
\begin{equation}
    C = \begin{pmatrix}
        0 & \mathds{1} \\
        \mathds{1} & 0\\
    \end{pmatrix}.
\end{equation}
This allows us to transform Hamiltonian into a basis in which it is antisymmetric at the high-symmetry points, $k_3=0, \pi$. This unitary transformation is given by:
\begin{equation}
    \tilde{\mathcal{H}}^{\rm NW}_{\rm sc}(k_3)  = U^{\dagger}\mathcal{H}^{\rm NW}_{\rm sc}(k_3) U, \qquad U = V\sqrt{\Lambda_C}^{-1},
    \label{hamasym}
\end{equation}
where where $V$ is the matrix of the eigenvectors of $C$ and $\Lambda_C$ is a diagonal matrix made up of eigenvalues of $C$.

For an antisymmetric matrix we can define a polynomial called Pfaffian which squares to the determinant of that matrix:
\begin{equation}
    \mathrm{Pf}(A) = \frac{1}{2^n n!}\sum_{\sigma \in S_{2n}} \mathrm{sign} (\sigma) \prod_i^n A_{\sigma (2i-1), \sigma(2i)},
\end{equation}
where $S_{2n}$ is the group of permutations of sets with $2n$ elements. This quantity is useful to construct an invariant of particle-hole symmetric systems. The Pfaffian of an antisymmetric Hamiltonian can change sign if one of the eigenvalues passes through zero and changes sign. Eigenvalues of particle-hole symmetric Hamiltonian always come in pairs $E_n$ and $-E_n$, so the change of the sign can happen only when there is a gap closing, which is an indication of a topological phase transition. Indeed, we can construct a $\mathbb{Z}_2$ topological invariant using Pfaffian signs of the antisymmetric Hamiltonian at the high symmetry points:
\begin{equation}\label{eq:inv}
    \nu_{\rm sc} = \frac{1}{2} - \frac{1}{2}\mathrm{sign} \left\{ \mathrm{Pf} \left[i\tilde{\mathcal{H}}^{\rm NW}_{\rm sc}(0)\right]\mathrm{Pf}\left[i\tilde{\mathcal{H}}^{\rm NW}_{\rm sc}(\pi)\right]\right\},
\end{equation}
where $\tilde{\mathcal{H}}^{\rm NW}_{\rm sc}(k_3)$ is given by Eq. (\ref{hamasym}). From hermiticity of the Hamiltonian it follows that the antisymmetric matrices in the above formula are real. 

\begin{figure}[h]
\includegraphics[width=0.9\columnwidth]{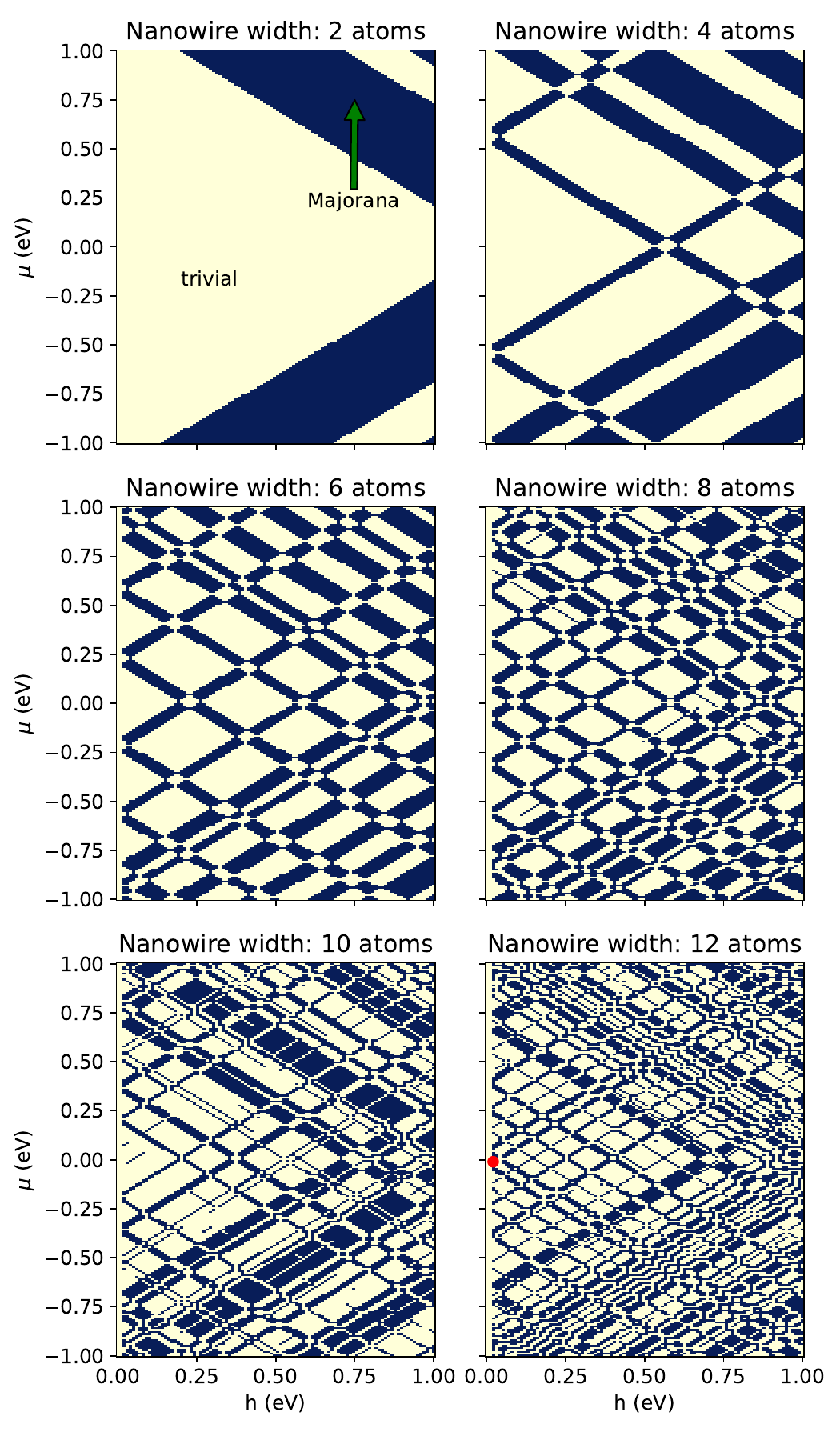}
\caption{\label{fig:scdiag} Phase diagrams of the $\mathbb{Z}_2$ invariant of the superconducting phase parametrized with chemical potential, $\mu$ and magnetic field magnitude, $B_z$. Light regions indicate the trivial phase and dark regions, a phase with Majorana end states. 
}
\end{figure}

To compute the $\nu_{\rm sc}$ invariant, we use algorithm proposed in \cite{Wimmer11} that utilizes Schur's decomposition of real antisymmetric matrices. Fig. \ref{fig:scdiag} shows phase diagrams of trivial and topological regions for six NW widths. The interesting observation is that for increasing thickness of the systems the non-trivial regions in the diagram form denser and denser mesh. This is similar to the results obtained for $[001]$ NWs \cite{Brzezicki_snte_nw}, but here the situation is more favorable for a non-trivial phase because it starts for smaller fields and chemical potential for comparable width of a NW.

\begin{figure}[h]
\includegraphics[width=0.9\columnwidth]{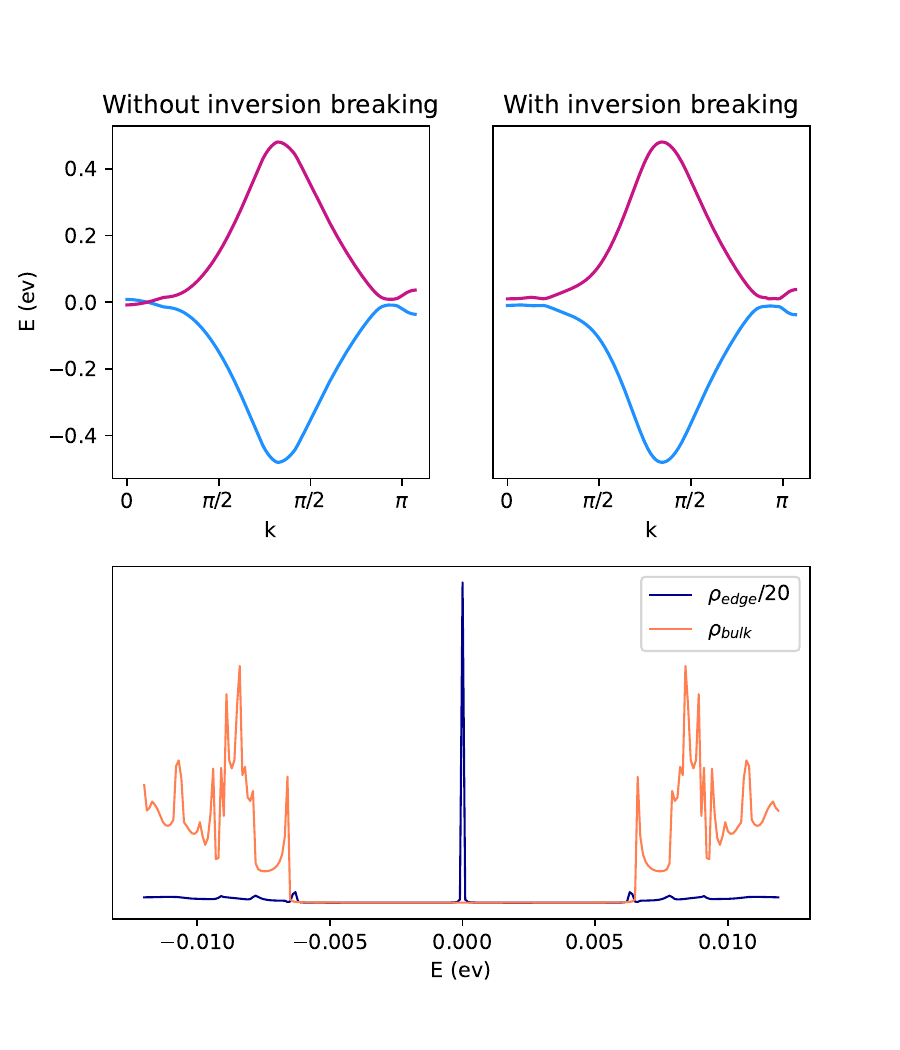}
\caption{\label{fig:majorana} Results for the system in non-trivial phase with parameters: $\mu = -0.01$ eV, $h = 0.02$ eV, $\mathrm{thickness} =$ 12 atoms,  (marked by red dot in Fig. \ref{fig:scdiag}). \textbf{Top} row: bands of the Hamiltonian without inversion breaking field (left, gapless) and with inversion breaking field, $\boldsymbol{\lambda}_R = (0.12, 0.14, 0.14)$ (right, fully gapped). \textbf{Bottom} row:
density of states in the bulk, $\rho_{bulk}$, and at the ends of the wire, $\rho_{edge}$.
}
\end{figure}

The invariant $\nu_{\rm sc}$ is a bulk invariant computed for system with periodic boundary conditions. To observe edge states it is necessary to open the system and see what happens at the ends of the wire. For Majorana end states, we also need a gapful system. To achieve it, we can add an inversion breaking field in the form of Rashba coupling term to the Hamiltonian: 
\begin{equation}
{\cal H}_R(\mathbf{k})=\boldsymbol{\lambda}_R\cdot\sin\mathbf{k}\times\boldsymbol{\sigma}.
\end{equation}
Now we can choose parameters of the model in the non-trivial phase marked by red dot on Fig. \ref{fig:scdiag} and look at the bulk gap and the end-states, see Fig. \ref{fig:majorana}. We see that without Rashba term the gap closes near high-symmetry points: we get bulk Majorana zero-energy states. 
On the other hand including ${\cal H}_R$ gives a fully gapped spectrum
with gap of the order of $10^{-1}$ eV. To confirm the presence of the Majorana end-states we have calculated bulk and edge densities of states using the same iterative Green's function algorithm as in the normal state. In the bottom plot of Fig. \ref{fig:majorana} we can clearly see tall peak at zero energy in the edge density of states confirming the presence of Majorana end-states.

\section{Discussion and conclusions}

We have shown that $[110]$ SnTe nanowire support end states in the normal state related to the non-trivial $M\mathbb{Z}$ mirror invariant. These non-trivial insulating states are triggered by non-zero Zeeman magnetic field in the direction parallel to the wire. In between different gapped $M\mathbb{Z}$-phases we also find Weyl semi-metal phases protected by the screw-axis symmetry. This is indeed quite similar to the phase diagram of the $[001]$ SnTe nanowire studied earlier \cite{Brzezicki_snte_nw}. However, there are important differences: the screw axis symmetry is twofold in the present nanowire and, more importantly, the gap-inverted insulating phases are protected by an exact topological invariant. Such invariant was absent in the case of   $[001]$ nanowire and one could only rely in pseudospin textures to describe gap inversions. 

For practical reasons we have concentrated on relatively thin nanowires and have found that the phase diagram as function of thickness and Zeeman magnetic field is quite complicated. Nevertheless, from the general trends in the thickness dependence we argue that for realistic nanowire thicknesses the topologically nontrivial phases can be reached with experimentally feasible values of the Zeeman field.

Finally, we have found that the superconducting SnTe nanowires support gapless bulk Majorana modes in the presence of inversion symmetry, and by introducing inversion-symmetry-breaking field, the bulk Majorana modes become gapped and topologically protected localized Majorana zero modes appear at the ends of the wire.
This finding opens up possibilities to control and create Majorana zero modes by controlling the inversion-symmetry breaking fields. Quite interestingly, Majorana end-states seem to appear for smaller chemical potential and Zeeman magnetic field than for $[001]$ nanowires which suggest that $[110]$ growth direction is more favorable from the point of view of search for Majoranas.

There exists various possibilities to experimentally probe 
end states in the normal and superconducting phases of the nanowires. High-quality transport studies are definitely the best way to study these systems. Ideally, the SnTe bulk materials would be insulators where the Fermi level is inside the insulating gap. The interesting physics, including the topological surface states, appear in this range of energies. Unfortunately, in reality the SnTe bulk materials typically have a large residual carrier density due to defects, which poses a significant obstacle for the studies of topological transport effects. Therefore it is of crucial importance to improve the control of the carrier density in SnTe materials. In comparison to the bulk systems the nanowires have the advantage that the carrier density can be more efficiently controlled with gate voltages. 
Tunneling measurements are possible also in the presence of a large carrier density because one can probe the local density of states as a function of energy by voltage-biasing the tip.
One may also try to observe the end-states using nano-ARPES but obtaining simultaneously both high spatial and energy resolution may be a difficult challenge.  
The topologically protected gapless Majorana bulk modes could be probed via thermal conductance measurements, and they may also be detectable by measuring electrical shot-noise power or magnetoconductance oscillations in a ring geometry \cite{Akh11}. The Majorana zero modes give rise to various effects, such as a robust zero-bias peak in the conductance \cite{Law09} and $4\pi$ Josephson effect \cite{Kit01,Lut10}, but the ultimate goal in the physics of Majorana zero modes is of course to observe effects directly related to the  non-Abelian Majorana statistics \cite{Sar15, Beenakker20, flensberg2021engineered}.
The Majorana zero modes can be realized even if a significant residual carrier density is present as illustrated in our phase diagrams. However, the new experimental challenge in this case is that the topologically nontrivial phase becomes more and more fragmented in thick wires.

\section*{Acknowledgements}
A.K and W.B. acknowledge support by Narodowe Centrum Nauki (NCN, National Science Centre, Poland) Project No. 2019/34/E/ST3/00404. W.B acknowledge partial support by the Foundation for Polish Science through the IRA Programme co-financed by EU within SG OP.

\appendix 

\section{Construction of the nanowire Hamiltonian and the symmetry operators \label{app:nanowire}}

In this section we give explicit expressions for the different symmetry operators of the nanowire Hamiltonian. Our starting point is the bulk Hamiltonian (\ref{eq:ham}). The nearest-neighbor hopping matrices are
\begin{eqnarray}
h_x &=& (1+\cos k_1)\sigma_x \otimes \sigma_x +(\sin k_1) \sigma_y \otimes \sigma_x\nonumber\\
h_y &=& (1+\cos k_2)\sigma_0 \otimes \sigma_x +(\sin k_1) \sigma_z \otimes \sigma_y\nonumber\\
h_z &=& (\cos (k_1-k_3)+\cos k_3)\sigma_x \otimes \sigma_x \nonumber\\
&+&(\sin (k_1-k_3)+ \sin k_3) \sigma_y \otimes \sigma_x
\end{eqnarray}
and the next-nearest-neighbor hopping matrices are
\begin{eqnarray}
2h_{xy} &=& (1+\cos k_1 + \cos k_2 + \cos(k_1+k_2))\sigma_x \otimes \sigma_0\nonumber\\
&+&(1-\cos k_1 - \cos k_2 + \cos(k_1+k_2))\sigma_x \otimes \sigma_z\nonumber\\
&+&(\sin k_1 - \sin k_2 + \sin(k_1+k_2))\sigma_y \otimes \sigma_0\nonumber\\
&+&(-\sin k_1 + \sin k_2 + \sin(k_1+k_2))\sigma_y \otimes \sigma_z\nonumber \\
\end{eqnarray}
\begin{eqnarray}
2h_{yx} &=& (1+\cos k_1 + \cos k_2 + \cos(k_1-k_2))\sigma_x \otimes \sigma_0 \nonumber\\
&+&(-1+\cos k_1 + \cos k_2 - \cos(k_1-k_2))\sigma_x \otimes \sigma_z \nonumber\\
&+&(\sin k_1 + \sin k_2 + \sin(k_1-k_2))\sigma_y \otimes \sigma_0\nonumber\\
&+&(\sin k_1 + \sin k_2 - \sin(k_1+k_2))\sigma_y \otimes \sigma_z
\end{eqnarray}
\begin{eqnarray}
2h_{yz} &=& (\cos k_3\!+\!\cos (k_2\!+\!k_3)\!\nonumber\\
&&\! +\cos (k_1\!-\!k_3)\!+\!\cos (k_1\!-\!k_2\!-\!k_3))\sigma_x \otimes \sigma_0\nonumber\\
&+& (-\cos k_3\!+\!\cos (k_2\!+\!k_3)\!+\!\nonumber\\
&&\cos (k_1\!-\!k_3)\!-\!\cos (k_1\!-\!k_2\!-\!k_3))\sigma_x \otimes \sigma_z\nonumber\\
&+& (\sin k_3\!+\!\sin (k_2\!+\!k_3)\!+\!\nonumber\\
&&\sin (k_1\!-\!k_3)\!+\!\sin (k_1\!-\!k_2\!-\!k_3))\sigma_y \otimes \sigma_0\nonumber\\
&+& (-\sin k_3\!+\!\sin (k_2\!+\!k_3)\!+\!\nonumber\\
&&\sin (k_1\!-\!k_3)\!-\!\sin (k_1\!-\!k_2\!-\!k_3))\sigma_y \otimes \sigma_z\nonumber\\
\end{eqnarray}
\begin{eqnarray}
2h_{zy} &=& (\cos k_3\!+\!\cos (k_2\!-\!k_3)\!\nonumber\\
&&+\!\cos (k_1\!-\!k_3)\!+\!\cos (k_1\!+\!k_2\!-\!k_3))\sigma_x \otimes \sigma_0\nonumber\\
&+& (\cos k_3\!-\!\cos (k_2\!-\!k_3)\!\nonumber\\
&&-\!\cos (k_1\!-\!k_3)\!+\!\cos (k_1\!+\!k_2\!-\!k_3))\sigma_x \otimes \sigma_z\nonumber\\
&+& (\sin k_3\!-\!\sin (k_2\!-\!k_3)\!\nonumber\\
&&+\!\sin (k_1\!-\!k_3)\!+\!\sin (k_1\!+\!k_2\!-\!k_3))\sigma_y \otimes \sigma_0\nonumber\\
&+& (\sin k_3\!+\!\sin (k_2\!-\!k_3)\!\nonumber\\
&&-\!\sin (k_1\!-\!k_3)\!+\!\sin (k_1\!+\!k_2\!-\!k_3))\sigma_y \otimes \sigma_z\nonumber\\
\end{eqnarray}
\begin{eqnarray}
h_{xz} &=& 2(\cos k_3) \sigma_0 \otimes \sigma_0\nonumber\\
h_{zx} &=& 2(\cos (k1-k_3))\sigma_0 \otimes \sigma_0,\nonumber\\
\end{eqnarray}
where $\sigma_x, \sigma_y, \sigma_z$ are Pauli matrices and $\sigma_0$ is 2$\times$2 identity matrix. Note that they are not related to the spin of electrons but describe four internal sites of a unit cell.
The matrix $\Sigma$ related to the sublattice degree of freedom is given by:
\begin{equation}
    \Sigma = - \sigma_0 \otimes \sigma_z.
\end{equation}

To obtain the lower dimensional Hamiltonians, we can first expand the Hamiltonian in Fourier components as
\begin{equation}
\mathcal{H} \left(\textbf{k} \right) =
\sum_{j_1=-r_1}^{r_1}\sum_{j_2=-r_2}^{r_2}
e^{i(j_1k_1+j_2k_2)}H_{j_1,j_2}(k_3),
\end{equation}
where $r_i$ stands for maximum hopping ranges (in multiples of unit cells) along lattice vectors $\textbf{a}_i$.
Then the Hamiltonian of the nanowire of the thickness $N_x\times N_y$
unit cells can be obtained from the bulk one by a formula:
\begin{equation}
\mathcal{H}^{\rm NW}(k_3) = \sum_{j_1=-r_1}^{r_1}\sum_{j_2=-r_2}^{r_2} T_{N_x}^{j_1}\otimes T_{N_y}^{j_2}\otimes H_{j_1,j_2}(k_3)
\end{equation}
with $T_{N_{\alpha}}$ being $N_{\alpha} \times N_{\alpha}$ matrix of the form 
\begin{equation}
T_{N_{\alpha}}=
\begin{pmatrix}0 & 1\\
 & 0 & 1\\
 &  & \ddots & \ddots\\
 &  &  & 0 & 1\\
 &  &  &  & 0
\end{pmatrix},\quad
T_{N_{\alpha}}^{-1}:=T_{N_{\alpha}}^T.
\end{equation}
The nanowire has a two-fold
screw-axis symmetry, which is described by a unitary operator:
\begin{eqnarray}
S_c\left(k_{3} \right) &=& ie^{-i(N_x-1)k_3}\times P_{\rm inter}(k_3)\otimes \exp\left[i \tfrac{\pi}{2\sqrt 2}(\sigma_{x}+\sigma_{z})\right] \nonumber\\
&\otimes& \exp\left[i \tfrac{\pi}{\sqrt 2}(L_{x}+L_z)\right]\otimes  P_{\rm intra} \left( k_3 \right). \label{eqSc} 
\end{eqnarray} 
with the intra-cell matrix, acting on four internal sites of the unit cell:
\begin{equation}
P_{\rm intra} \left( k_3 \right) = 
\cos\tfrac{k_3}{2}\sigma_x\otimes\sigma_0 +
\sin\tfrac{k_3}{2}\sigma_y\otimes\sigma_0
\end{equation}
and the inter-cell one, acting on the $N_x\times N_y$  lattice of the unit cells:
\begin{eqnarray}
P_{\rm inter} \left( k_3 \right)=\hspace{6 cm}\\ 
\begin{small}\begin{pmatrix}
0 & \cdots & \cdots & 0& 1 \\
0 & \cdots & \cdots & e^{i2k_3} & 0 \\
0 & \cdots & e^{i4k_3}  & 0 & 0\\
\vdots & \iddots & \vdots  & \vdots & \vdots\\
e^{i2(N_x-1)k_{3}} & \cdots & 0 & 0 & 0
\end{pmatrix}\end{small}
\otimes
\begin{small}\begin{pmatrix}
0 & \cdots & \cdots & 0& 1 \\
0 & \cdots & \cdots & 1 & 0 \\
0 & \cdots & 1  & 0 & 0\\
\vdots & \iddots & \vdots  & \vdots & \vdots\\
1 & \cdots & 0 & 0 & 0
\end{pmatrix}\end{small}.
\nonumber
\end{eqnarray}
We find that:
\begin{eqnarray}
[\mathcal{H}^{\rm NW}(k_3),S_c\left(k_{3} \right)]&\equiv& 0\\
S_c\left(k_{3} \right)^2&\equiv& \mathbbm{1}\\
\Tr S_c\left(k_{3} \right)&\equiv& 0\nonumber
\end{eqnarray}
Another important symmetry is the inversion described by a unitary operator:
\begin{equation}
I = P_{\rm inter}(0)\otimes \mathbbm{1}_2
\otimes \mathbbm{1}_3\otimes  P_{\rm intra}(0).  
\end{equation} 
We find that
\begin{eqnarray}
I\mathcal{H}^{\rm NW}(k_3)I^{-1}&\equiv& \mathcal{H}^{\rm NW}(-k_3),\\
IS_c(k_3)I^{-1}&\equiv& S_c(-k_3).
\end{eqnarray}
Finally, at zero Zeeman field $h$ the system is 
time-reversal symmetric with the time-reversal operator
taking a standard spinful form of
\begin{equation}
{\cal T} = {\cal K}\,\mathbbm{1}_{\!N_x}\otimes\mathbbm{1}_{\!N_y}\otimes \sigma_y
\otimes \mathbbm{1}_3\otimes  \mathbbm{1}_4,
\end{equation} 
with $\cal K$ being complex conjugation and having standard relation with the Hamiltonian:
\begin{eqnarray}
{\cal T}\mathcal{H}^{\rm NW}(k_3){\cal T}^{-1}&\equiv& \mathcal{H}^{\rm NW}(-k_3).
\end{eqnarray}
Since ${\cal T}^2=-1$ the Hamiltonian has Kramers-degenerate spectrum for $k_3=0,\pi$ when $h=0$. Due to presence of the inversion symmetry the Kramers degeneracy is present also at any other $k$-point when $h=0$. The relation between time-reversal and the screw-axis is the following:
\begin{eqnarray}
{\cal T}S_c(k_3){\cal T}^{-1}&\equiv& -S_c(-k_3).
\end{eqnarray}
Consequently it mean that in the eigenbasis of the screw-axis operator the Hamiltonian and its symmetries take a block form of:
\begin{eqnarray}
\tilde{\mathcal{H}}^{\rm NW}(k_3) &=& \mathcal{H}_+(k_3)\oplus\mathcal{H}_-(k_3) \\
\tilde{I} &=& I_+\oplus I_-
\end{eqnarray}
and 
\begin{eqnarray}
\tilde{{\cal T}}={\cal K}\begin{pmatrix}0 & V\\
V^{\dagger} & 0
\end{pmatrix}.
\end{eqnarray}
Note that the exact form of the above matrix blocks depends on the choice of gauge of the eigenvectors of $S_c(k_3)$ which are non-trivial functions of $k_3$. With a proper choice of gauge blocks $I_{\pm}$ and $V$ can be found indeed $k_3$-independent. However, due to the nonsymmorphic character of the screw-axis symmetry we find that the Hamiltonian in the new basis becomes $4\pi$-periodic, i.e.,
\begin{equation}
\mathcal{H}_{\pm}(k_3) = \mathcal{H}_{\pm}(k_3+4\pi)\not=
\mathcal{H}_{\pm}(k_3+2\pi),
\end{equation}
and for $h\not=0$ the eigenvalues of $\mathcal{H}_{\pm}(k_3)$ and $\mathcal{H}_{\pm}(k_3+2\pi)$ are different.
On the other hand, for $h=0$ the two blocks, $\mathcal{H}_{+}(k_3)$ and $\mathcal{H}_{-}(k_3)$, are time-reversal partners and share the same eigenvalues. This means that even for $h=0$ the time-reversal symmetry is broken inside each block. On the other hand for any $h$ both blocks are inversion-symmetric, obeying:
\begin{eqnarray}
I_{\pm}\mathcal{H}_{\pm}(k_3)I_{\pm}^{-1}&\equiv& \mathcal{H}_{\pm}(-k_3).
\end{eqnarray}
Therefore from the point of view of inversion in this basis we have two distinct high-symmetry points $k_3=0$ and $k_3=2\pi$ for which 
\begin{eqnarray}
[I_{\pm},\mathcal{H}_{\pm}(0)]=[I_{\pm},\mathcal{H}_{\pm}(2\pi)].
\end{eqnarray}

\section{Mirror symmetry and the choice of unit cell \label{app:mirror}}

\begin{figure}[t!]
\includegraphics[width=0.8\columnwidth]{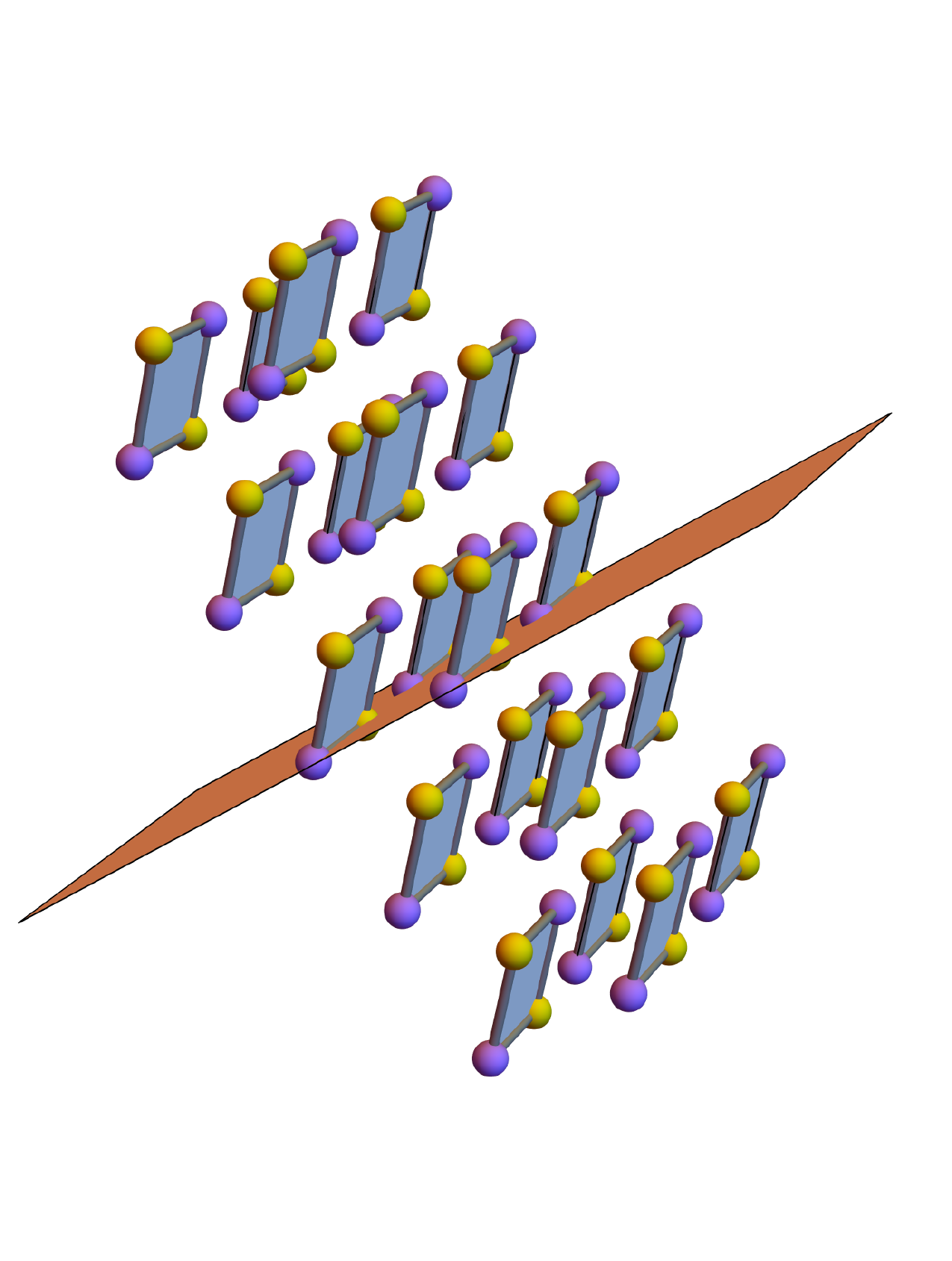}
\caption{\label{fig:sym_plane} Model of the system with mirror symmetry compatible ends.}
\end{figure}

The $[101]$-mirror reflection symmetry of the NW is given by:
\begin{eqnarray}
M_r\left(k_{3} \right) &=& i\,G_{\rm inter}(k_3)\otimes \mathbbm{1}_{N_y} \otimes\exp\left[i \tfrac{\pi}{2\sqrt 2}(\sigma_{x}+\sigma_{z})\right] \nonumber\\
&\otimes& \exp\left[i \tfrac{\pi}{\sqrt 2}(L_{x}+L_z)\right]\otimes  G_{\rm intra} \left( k_3 \right).
\end{eqnarray} 
with $G$ being diagonal inter- and intra-cell gauge matrix
given by
\begin{eqnarray}
G_{\rm inter}\left( k_3 \right) &=& {\rm diag} \left(1,e^{-i2k_3},\dots,e^{-i2(N_x-1)k_3} \right),\\
G_{\rm intra}\left( k_3 \right) &=& {\rm diag} \left(1,1,e^{-ik_3},e^{-ik_3} \right).
\end{eqnarray}
The prefactor is chosen such that $M_r(k_3)$ satisfies $M_r(0)^2=M_r(\pi)^2=1$. 
It is possible to define a gauge transformation
\begin{equation}
\Gamma\left(k_{3} \right) = {\rm diag} \left(1,e^{k_3},\dots,e^{i(N_x-1)k_3} \right)
\otimes\mathbbm{1}_{24N_y},
\end{equation} 
such that in the new basis,
\begin{eqnarray}
\tilde{\mathcal{H}}^{\rm NW}(k_3) &=&\Gamma\left(k_{3} \right)\mathcal{H}^{\rm NW}(k_3)\Gamma\left(-k_{3} \right), \\
\tilde{M}_r(k_3) &=&\Gamma\left(k_{3} \right)M_r(k_3)\Gamma\left(k_{3} \right), 
\end{eqnarray}
mirror symmetry take a simpler form of 
\begin{eqnarray}
\tilde{M}_r(k_3) &=& i\,\mathbbm{1}_{N_x}\otimes \mathbbm{1}_{N_y} \otimes\exp\left[i \tfrac{\pi}{2\sqrt 2}(\sigma_{x}+\sigma_{z})\right] \nonumber\\
&\otimes& \exp\left[i \tfrac{\pi}{\sqrt 2}(L_{x}+L_z)\right]\otimes  G_{\rm intra} \left( k_3 \right).
\end{eqnarray} 
Physically this means choosing a unit cell of the NW as the one show in Fig. \ref{fig:sym_plane}.

\section{Gap diagrams versus magnetization steps \label{app:gap}}
\begin{figure}[h]
\includegraphics[width=0.99\columnwidth]{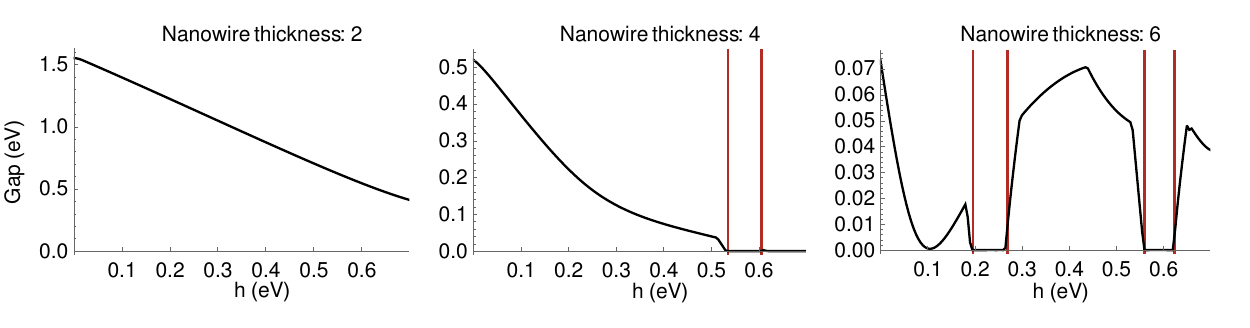}
\includegraphics[width=0.99\columnwidth]{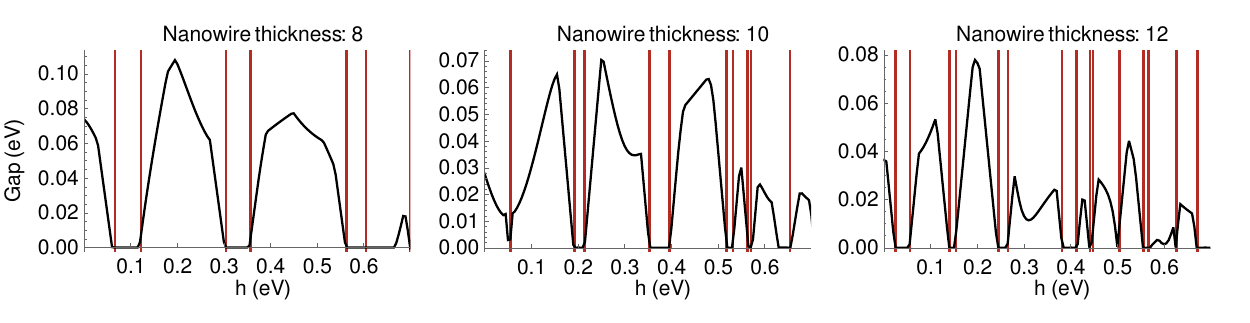}
\caption{\label{fig:gaps} Energy gap widths of nanowires with varying thickness (number of atoms). Red gridlines indicate magnetization steps from Figure \ref{fig:magn}. }
\end{figure}

In Fig. \ref{fig:gaps} we show the energy gap curves for increasing NW thickness vs Zeeman magnetic field.

\section{Pseudospin textures \label{app:stext}}

\begin{figure}[t!]
\includegraphics[width=0.49\columnwidth]{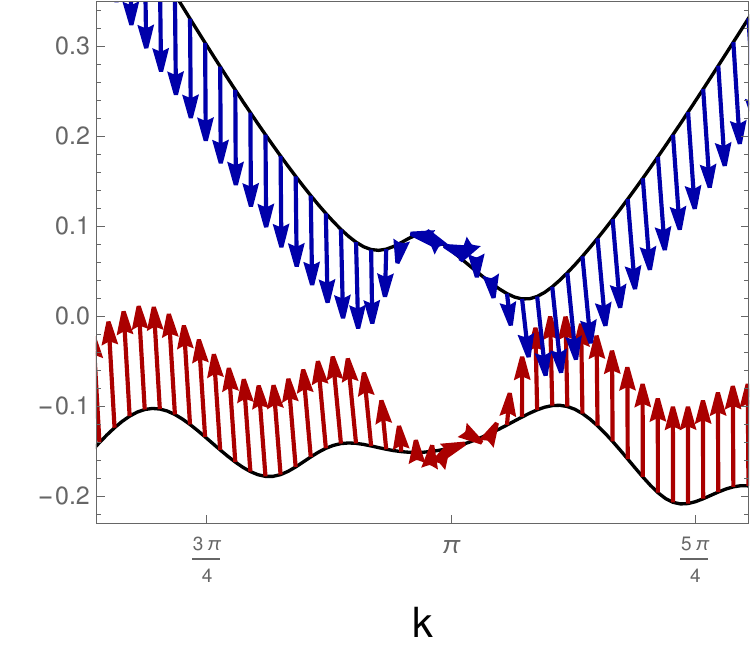}
\includegraphics[width=0.49\columnwidth]{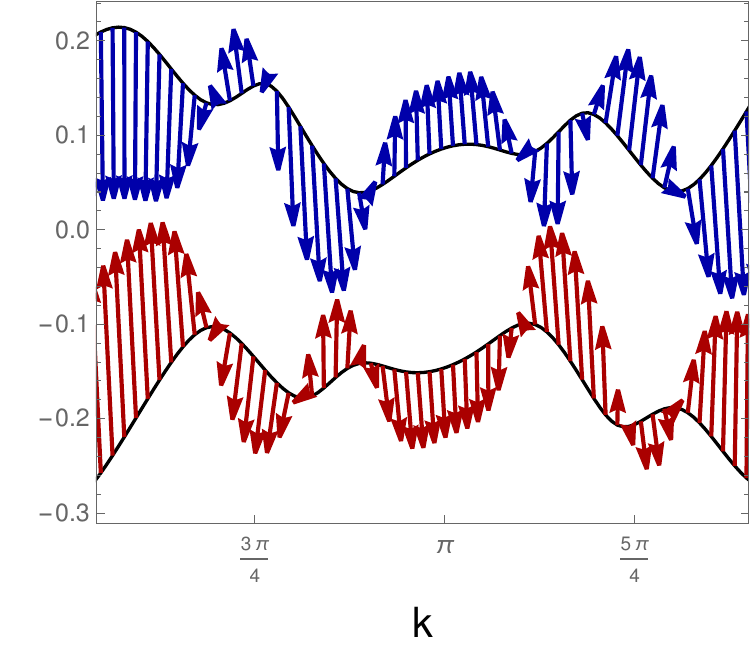}
\caption{\label{fig:spin}Spin structures of simple (left image) and gap-inverted (right image) insulating nanowires 3 unit cells thick.}
\end{figure}

We define a sublattice pseudospin of a given band as
\begin{eqnarray}
s_n^{\alpha}(k_3) &=& \langle E_n(k_3)|\,\mathds{1}_{\!N_x}\otimes\mathds{1}_{\!N_y}\otimes
    \mathds{1}_2 \nonumber\\
    &\otimes &\mathds{1}_3\otimes(\sigma_0 \otimes\sigma_{\alpha}) |E_n(k_3) \rangle,
\end{eqnarray}
with $\alpha=x,y,z$. The example of such a textures for two bands around the gap for a $3$-cell wide NW are shown in Fig. \ref{fig:spin}. We see that the texture gets a spin-flop in the topologically non-trivial phase. Note that
only $x$ and $z$ components are shown because $s_n^{z}$ is 
around factor of $100$ smaller.

\clearpage
\eject
\bibliography{apssamp} 

\end{document}